\documentclass[a4paper]{jpconf}
\usepackage{graphicx}
\usepackage{amsmath,amssymb}

\newcommand{\ket}[1]{| {#1} \rangle}

\def\vect#1{{\mbox{\boldmath $#1$}}}

\begin{document}
\title{Systematic investigation of $E1$ strength\\ for the isotopes from $Z$ = 28 to 50}

\author{Shuichiro Ebata$^{1,2}$,
Takashi Nakatsukasa$^{2,3}$
and Tsunenori Inakura$^{2}$}

\address{$^1$ Center for Nuclear Study, University of Tokyo, Wako-shi 351-0198, Japan}
\address{$^2$ RIKEN Nishina Center, Wako-shi 351-0198, Japan}
\address{$^3$ Center for Computational Sciences, University of Tsukuba,
Tsukuba 305-8571, Japan}

\ead{ebata@cns.s.u-tokyo.ac.jp}

\begin{abstract}
We carry out a systematic study of electric dipole mode ($E1$) 
for neutron-rich isotopes from nickel ($Z$=28) to tin ($Z$=50) using 
a time-dependent mean-field theory. 
Our time-dependent scheme is 
the canonical-basis time-dependent Hartree-Fock-Bogoliubov theory 
which can self-consistently describe nuclear dynamics with pairing correlation.
We focus our discussion on the pygmy dipole resonance (PDR) and $E1$ polarizability ($\alpha_{\rm D}$). 
The correlation between neutron-skin thickness and PDR strongly depends on 
the neutron number, but 
the correlation between the skin thickness and $\alpha_{\rm D}$ 
is much more stable. 
\end{abstract}

\section{Introduction}

Recent progress in radioactive isotope facilities and
experimental techniques allows us to generate and 
to study neutron-rich nuclei. 
In the neutron-rich region far from stability line, 
the neutron-skin and halo structures 
may appear in the ground state \cite{1}. 
Due to their exotic structures in the ground state, 
new elementary modes are expected. 
The pygmy dipole resonance (PDR) is known as a candidate of 
such a characteristic excited state in the neutron-rich nuclei, 
which appears in low energy region, sometimes interpreted 
as a neutron-skin mode. 
Previous studies indicate that there is the linear relation 
between PDR and neutron-skin thickness \cite{2,3}. 
On the other hand, Ref.\cite{4} indicates that  
the correlation between PDR and the skin is small 
from the covariance analysis, and that the electric dipole ($E1$) 
polarizability is much more correlated to the neutron skin than PDR \cite{4}. 
Recently, the dipole polarizability has been measured for $^{206}$Pb at 
RCNP, Osaka University \cite{5}. 
 
In the present paper, the relation among neutron-skin, PDR and $E1$ polarizability, 
will be shown for over 350 isotopes, 
which is obtained by the linear response calculation using the time-dependent scheme. 

\section{Formulation}

In order to study the dynamics of many nuclei including superfluid deformed nuclei, 
we proposed the canonical-basis time-dependent Hartree-Fock-Bogoliubov (Cb-TDHFB) theory. 
The Cb-TDHFB can be derived from the full TDHFB equation 
represented in the canonical basis and on the assumption 
that the pair potential can be approximated to be diagonal in the canonical basis \cite{6}.

Introducing the time-dependent canonical states $\ket{\phi_k(t)}$ and
$\ket{\phi_{\bar k}(t)}$,
we express the TDHFB state in the canonical (BCS) form as
\begin{equation}
\ket{\Psi(t)}=\prod_{k>0} \left\{
u_k(t) + v_k(t) c_k^\dagger(t) c_{\bar k}^\dagger(t) \right\} \ket{0} ,
\end{equation}
where $u_k(t), v_k(t)$ are time-dependent BCS factors. 
It should be note that the pair of states, $k$ and $\bar k$,
are not necessarily related to each other by the time-reversal,
$\ket{\phi_{\bar{k}}(t)}\neq T\ket{\phi_{k}(t)}$.
The diagonal approximation of the pair potential leads to the following equations:
\begin{subequations}
\label{Cb-TDHFB}
\begin{eqnarray}
\label{dphi_dt}
&&i\frac{\partial}{\partial t} \ket{\phi_k(t)} =
(h(t)-\eta_k(t))\ket{\phi_k(t)} , \quad\quad
i\frac{\partial}{\partial t} \ket{\phi_{\bar k}(t)} =
(h(t)-\eta_{\bar k}(t))\ket{\phi_{\bar k}(t)} , \\
\label{drho_dt}
&&
i\frac{d}{dt}\rho_k(t) =
\kappa_k(t) \Delta_k^{\ast}(t)
-\kappa_k^{\ast}(t) \Delta_k(t) , \\
\label{dkappa_dt}
&&
i\frac{d}{dt}\kappa_k(t) =
\left(
\eta_k(t)+\eta_{\bar k}(t)
\right) \kappa_k(t) +
\Delta_k(t) \left( 2\rho_k(t) -1 \right) .
\end{eqnarray}
\end{subequations} 
These basic equations determine the time evolution of
the canonical states, $\ket{\phi_k(t)}$ and $\ket{\phi_{\bar k}(t)}$,
their occupation $\rho_k(t)\!=\!|v_{k}(t)|^{2}$, and 
pair probabilities $\kappa_k(t)\!=\!u_{k}(t)v_{k}(t)$. 
$\Delta_k(t)$ is the gap energy, 
\begin{eqnarray}
 \Delta_k(t) = \sum_{l>0} G f(\varepsilon_{k}^{0}) f(\varepsilon_{l}^{0}) \kappa_{l}(t), 
\end{eqnarray}
where $G$ is a the pairing strength determined by the smoothed pairing method \cite{7}, and 
$f(\varepsilon)$ is a cutoff function which defines the number of canonical basis \cite{6}. 
The cutoff function is time-independent with the single-particle energies $\varepsilon_{k}^{0}$ at the ground state. 

We solve the Cb-TDHFB equations in real time and calculate the linear response of the nucleus.
The numerical procedure of linear response is same as that of Ref.\cite{6}, 
we add an external field $\hat{V}_{\rm ext}(\vect{r},t)= -k\hat{F}_{E1}\delta(t)$ 
to the ground state of the nucleus, where $\hat{F}_{E1}=(N/A) \sum_p \hat{r}_p - (Z/A) \sum_n \hat{r}_n$, 
here $r=(x,y,z)$ and $k$ is an arbitrary small parameter. 
Then, we compute the time-evolution of the nuclear density using the time-dependent scheme. 
We obtain the strength function $S(E1;E)$ through the Fourier transformation of 
the time-dependent expectation value of $\hat{F}_{E1}$\cite{6}.
To quantify the PDR, we use the ratio:
\begin{eqnarray}
 \frac{m_{1}(E_{\rm c})}{m_{1}} \equiv \frac{\int^{E_{\rm c}} E\times S(E1; E) dE}{\int^{E_T} E\times S(E1; E) dE},
\end{eqnarray}
where we adopt the cutoff energy $E_{\rm c}$ = 10 MeV and 
the total energy $E_{T}$ = 100 MeV in the present calculation. 
The $E1$ polarizability $\alpha_{\rm D}$ is defined using $E1$ strength function as, 
\begin{eqnarray}
 \alpha_{\rm D} \equiv 2 \int^{E_T} \frac{S(E1; E)}{E} dE. 
\end{eqnarray} 

In this work, we use the three-dimensional Cartesian coordinate for the canonical states, 
$\phi_k (\vect{r},\sigma; t) = \langle \vect{r}, \sigma | \phi_k (t) \rangle$ 
with $\sigma = \pm 1/2$. 
The coordinate space is discretized in
the square mesh of $\Delta x$ = $\Delta y$ = $\Delta z$ = 1.0 fm 
in a sphere with radius of 15 fm.

\section{Results} 

Figure 1 shows the PDR ratio as a function of neutron-skin thickness 
which is defined by the difference between root mean square radii 
of neutrons and protons, from Ge to Sn isotopes.
In both panels, open circles and triangles indicate the isotopes 
with $N$=50 and 82, respectively. 
We can see the linear relation between the PDR ratio and 
the neutron-skin thickness for each isotopic chain with $N$=50 $\to$ 58. 
The slope of the linear correlation becomes small ass the proton number 
approaches $Z$=50. 
The slope for Sn isotope from $N$=50 to 58 
is about a half of that in Ge isotope. 
The PDR ratio strength jumps up again over $N$=82. 
The correlation between PDR and $N$-skin strongly depends on the neutron number. 
\begin{figure}[h]
  \begin{center}
   \includegraphics[keepaspectratio,width=60mm,angle=-90]{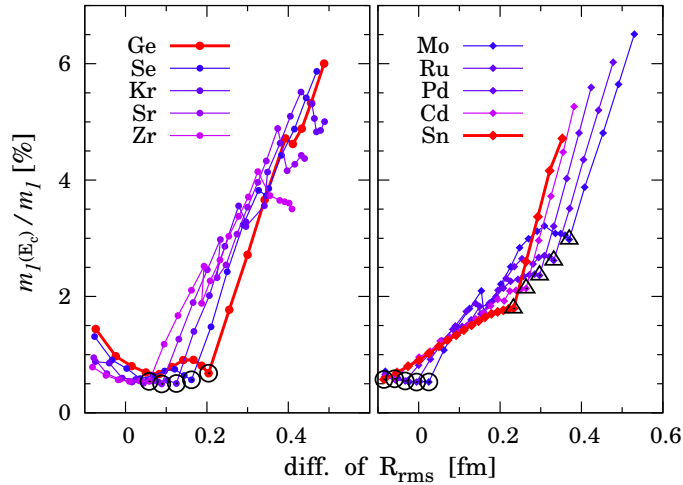}
  \end{center}
\caption{\footnotesize PDR ratio defined by Eq.(4),  
as a function of $N$-skin thickness from Ge to Sn isotopes. 
Open circle and triangle symbols indicate the isotope with $N$=50 and 82, respectively.}
\end{figure}

Figure 2 shows the $E1$ polarizability $\alpha_{\rm D}$ as a function of $N$-skin 
thickness for Sn isotopes. 
Circles (squares) are obtained with the SkM$^{\ast}$ (SkI3) Skyrme energy functional. 
Open symbols in Fig.2 are computed from the $E1$ strength excluding the PDR part; 
\begin{eqnarray}
 \tilde{\alpha}_{\rm D} \equiv 2 \int_{E_{\rm c}}^{E_T} \frac{S(E1; E)}{E} dE. 
\end{eqnarray}
This can be regarded as the dipole polarizability associated with the giant dipole resonance. 
The SkI3 parameters are made so as to reproduce the ordering of single-particle levels in Pb isotopes
obtained by the relativistic mean-field model. 
Especially, the part of the spin-orbit force is different from that of SkM$^\ast$. 
The polarizability $\tilde{\alpha}_{\rm D}$ is strongly correlated with the skin-thickness, 
although the contribution of the PDR becomes large in both results over $N$=82. 
In contrast to the relation between PDR and the skin, 
the correlation between $\alpha_{\rm D}$ and the skin thickness is 
more stable with respect to change of the neutron number and the energy functional. 
\begin{figure}[h]
  \begin{center}
   \includegraphics[keepaspectratio,width=60mm,angle=-90]{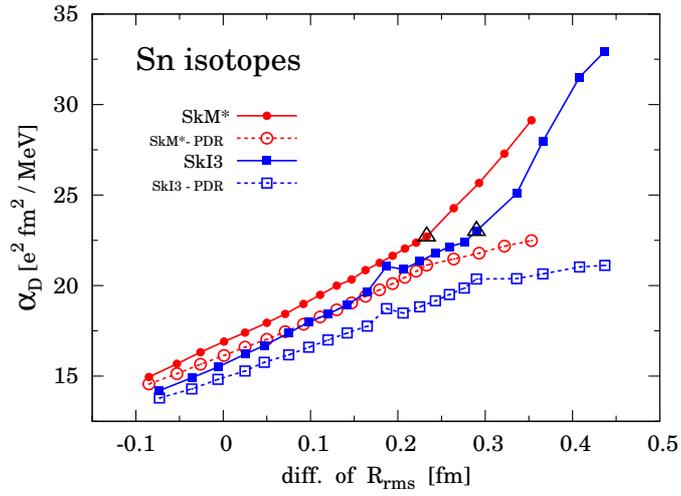}
  \end{center}
\caption{\footnotesize $E1$ polarizability $\alpha_{\rm D}$ defined by Eq.(5), 
as a function of neutron-skin thickness for Sn isotopes with $N=50-90$. 
Circles (squares) with solid lines indicate results with SkM$^{\ast}$ (SkI3) parameter set.
Open symbols with dashed lines are defined by Eq.(6)}
\end{figure}

In summary, we found the 
linear correlation between the PDR and the neutron-skin thickness, 
however, the correlation rather strongly depends on the neutron number. 
The correlation between $\alpha_{\rm D}$ and the neutron-skin thickness is 
more robust, especially if we remove the contribution from the PDR. 
This may suggest a better way to constrain the neutron-skin thickness by the 
observation of $E1$ strength function. 

\section*{Acknowledgments}
This work is supported by HPCI Strategic Program Field 5 hThe origin of matter and the universeh. 
The computational resources were provided by the RIKEN Integrated Cluster of Clusters (RICC), 
by the SR16000 at YITP in Kyoto University, and by the Joint Research Program 
at Center for Computational Sciences, University of Tsukuba.\\[5mm]
%%%%%%%%%%%%%%%%%%%%%%%%%%%%%%%%%%

\end{document}